\documentclass[floats,twocolumn,pra]{revtex4}
\usepackage{amssymb}
\usepackage{graphicx}
\usepackage{amsmath}
\usepackage{amsfonts}
\usepackage{accents}
\usepackage{color}
\newtheorem{theorem}{Theorem}

\newcommand{\SP}[1]{{{\textcolor{black}{#1}}}}

\begin{document}
\title{\SP{General upper bound for} conferencing keys \\in arbitrary quantum networks}
\author{Stefano Pirandola}
\affiliation{Department of Computer Science, University of York, York YO10 5GH, United Kingdom}

\begin{abstract}
Secure quantum conferencing refers to a protocol where a number of trusted
users generate exactly the same secret key to confidentially broadcast private
messages. By a modification of the techniques first introduced in     %simple
[Pirandola, arXiv:1601.00966], we derive a single-letter upper bound for the
maximal rates of secure conferencing in a quantum network with arbitrary
topology, where the users are allowed to perform the most powerful local
operations assisted by two-way classical communications, and the quantum
systems are routed according to the most efficient
multipath flooding strategies. More precisely, our analysis allows us to bound
the ultimate rates that are achievable by \SP{single-message} multiple-multicast
protocols, where $N$ senders distribute $N$ independent secret keys, and each
key is to be shared with an ensemble of $M$ receivers.

\end{abstract}
\maketitle

\section{Introduction}

Quantum information science~\cite{NiCh,first,HolevoBOOK,review,BraRMP} is
currently being developed at an unprecedented pace, with the field of quantum
key distribution (QKD)~\cite{BB84,Ekert,QKDadvance} already extended to \SP{quantum-secured networks~\cite{Frolic}} and even satellite communications~\SP{\cite{satellite1,satellite2}}.
Long-term plans to develop a fully-purpose quantum network, or `quantum
internet', are also contemplated from both a theoretical and experimental
point of view~\cite{Kimble,HybridINTERNET,Whener}. Building quantum networks
not only has the advantage of creating connectivity among many users, but also
gives the possibility to overcome the intrinsic fundamental limitation imposed
by the Pirandola-Laurenza-Ottaviani-Banchi (PLOB) bound~\cite{QKDpaper},
according to which the maximum number of quantum bits, entanglement bits
(ebits) or private/secret bits, that can be transmitted or generated at the
two ends of a lossy communication channel is limited to $-\log_{2}(1-\eta)$
bits per channel use, where $\eta$ is the channel's transmissivity. This limit
can be approached by point-to-point continuous variable protocols based on the
reverse coherent information~\cite{ReverseCAP,RevCohINFO} and can be beaten by
using suitably relay-assisted QKD protocols, such as the recently-introduced
twin-field QKD~\cite{Marco} \SP{(see also related experimental realizations~\cite{MarcoEXP,Liu})}, or by resorting to entanglement distillation
repeaters based on quantum memories~\cite{Briegel,Rep2,Rep3}.

Using techniques from classical network
theory~\cite{Slepian,Schrijver,Gamal,Cover,netflow} and tools more recently
developed in quantum information
theory~\cite{QKDpaper,Metro,nonPauli,TQC,BK2,Qmetro,revSENS},
Ref.~\cite{longVersion,netpaper} established tight bounds (and capacities) for
the repeater-assisted quantum communications over repeater chains and, more
general, network scenarios. These results were first developed for the unicast
case of a single sender and a single receiver in multi-hop quantum
networks, and then extended~\cite{longVersion,netpaper2} to multiend
configurations involving multiple senders and receivers, such as multiple
unicasts, multicasts, and multiple multicasts~\cite{Gamal}. All these
scenarios were considered in the setting of multiple independent messages, so
that each sender-receiver pair was assigned a different key with respect to
any other pair.

In this work, we extend the methodology of
Refs.~\cite{longVersion,netpaper,netpaper2} to the case of \SP{single-message multicasts, i.e., a scenario where} one or more senders aim to share exactly the same secret key with an ensemble of
receivers in a multi-hop quantum network. When the sender is only one, this
becomes a protocol of secure quantum conferencing in an arbitrary network
topology. \SP{Using tools of network simulation and stretching~\cite{longVersion}, we can write a general upper bound to the sum of all the key rates that the senders can optimally achieve in distributing their secret keys to the destination set of the receivers. This bound has a single-letter form in terms of the relative entropy of enetanglement (REE) and includes a minimization over suitable cuts of the network.}

%More precisely, our bounds are based on the techniques of quantum
%network simulation and the stretching of adaptive network protocols into a
%block form~\cite{longVersion}. In this way, we can exploits the tools of
%Ref.~\cite{QKDpaper} and write simple single-letter bounds.

It is important to
stress that this result not only applies to arbitrary network topologies but
also arbitrary dimensions of the Hilbert space, finite or
infinite. In other words, we consider quantum networks connected by
discrete-variable quantum channels, but also bosonic channels. Following the
methods in Refs.~\cite{QKDpaper,longVersion,netpaper,netpaper2,TQC,BK2}, we
can in fact introduce asymptotic notions of channel and network simulation
that allows us to rigorously prove results in the infinite-energy limit.

The paper is organized as follows. In Sec.~\ref{GeneralSECTION}\ we provide
preliminary notions for understanding the basic theory behind the next
derivation. In Sec.~\ref{MainSEC}\ we show our results for the distribution
of conferencing keys in a quantum network, extending the notion of
single-message multiple-multicast network to the quantum setting. Finally,
Sec.~\ref{SECconclusions}\ is for conclusions.

\section{Preliminaries\label{GeneralSECTION}}

%Let us start by providing a number of definitions and tools that are needed for our next derivation.

\subsection{Channel simulation}

Given a quantum channel $\mathcal{E}$, we can simulate it by means of local
operations (LOs) and classical communication (CC), briefly called LOCCs,
applied to the input state $\rho$ and a resource state $\sigma$. In other
words, we may write $\mathcal{E}(\rho)=\mathcal{T}(\rho\otimes\sigma)$. In
general, this simulation can be asymptotic, so that $\mathcal{E}(\rho
)=\lim_{\mu}\mathcal{T}^{\mu}(\rho\otimes\sigma^{\mu})$, for a sequence of
LOCCs $\mathcal{T}^{\mu}$ and resource states $\sigma^{\mu}$. Then, a channel
is called teleportation-covariant if it is covariant with respect to the
correction unitaries $U_{k}$ of teleportation~\cite{telereview}, i.e.,
finite-dimensional Pauli operators~\cite{teleBENNETT} or bosonic
displacements~\cite{Samtele,Samtele2}, depending on the dimension of the
Hilbert space. Channel $\mathcal{E}$ is teleportation-covariant if, for any
$U_{k}$, we have $\mathcal{E}(U_{k}\rho U_{k}^{\dagger})=V_{k}\mathcal{E}%
(\rho)V_{k}^{\dagger}$ for unitary $V_{k}$. In particular, for
$V_{k}=U_{k}$, $\mathcal{E}$ is called Weyl-covariant (or just Pauli covariant if the
dimension is finite). In discrete variables, for a tele-covariant
$\mathcal{E}$, we may write the simulation $\mathcal{E}(\rho)=\mathcal{T}%
_{\text{tele}}(\rho\otimes\sigma_{\mathcal{E}})$, where $\mathcal{T}%
_{\text{tele}}$ is teleportation and $\sigma_{\mathcal{E}}:=\mathcal{I}%
\otimes\mathcal{E}(\Phi)$ is the Choi matrix of the channel (here $\Phi$
denotes a finite-dimensional maximally-entangled state). In continuous
variables, we write $\mathcal{E}(\rho)=\lim_{\mu}\mathcal{T}_{\text{tele}%
}^{\mu}(\rho\otimes\sigma_{\mathcal{E}}^{\mu})$, where $\mathcal{T}%
_{\text{tele}}^{\mu}$ is the Braunstein-Kimble teleportation protocol based on
a two-mode squeezed vacuum (TMSV) state $\Phi^{\mu}$ with variance parameter
$\mu$, and $\sigma_{\mathcal{E}}^{\mu}:=\mathcal{I}\otimes\mathcal{E}%
(\Phi^{\mu})$ is a sequence of quasi-Choi matrices.

\subsection{Entanglement measures}

Given a state $\rho$, its REE~\cite{RMPrelent,VedFORMm,Pleniom} is defined as $E_{\mathrm{R}}%
(\rho)=\inf_{\gamma\in\text{\textrm{SEP}}}S(\rho||\gamma)$, where \textrm{SEP
}is the set of separable states and $S(\rho||\gamma):=\mathrm{Tr}\left[
\rho(\log_{2}\rho-\log_{2}\gamma)\right]  $ is the quantum relative entropy.
For an asymptotic state $\sigma:=\lim_{\mu}\sigma^{\mu}$ defined from a
sequence $\{\sigma^{\mu}\}$, we extend the definition considering
$E_{\text{\textrm{R}}}(\sigma)=\lim\inf_{\mu\rightarrow\infty}%
~E_{\text{\textrm{R}}}(\sigma^{\mu})$ (see Refs.~\cite{QKDpaper,TQC} for details). Typically, one identifies a suitable sequence of separable states
$\gamma^{\mu}$ and write the upper bound $E_{\text{\textrm{R}}}(\sigma
)\leq\lim\inf_{\mu\rightarrow\infty}S(\sigma^{\mu}||\gamma^{\mu})$. The
REE\ has important properties. First of all, it is monotonic under
trace-preserving LOCCs $\Lambda$, i.e., we have the data processing inequality
$E_{\text{\textrm{R}}}\left[  \Lambda\left(  \sigma\right)  \right]  \leq
E_{\text{\textrm{R}}}\left(  \sigma\right)  $. Second, it is sub-additive over
tensor products of states $\sigma^{\otimes n}$, i.e., we have $E_{\mathrm{R}%
}\left(  \sigma^{\otimes n}\right)  \leq nE_{\mathrm{R}}\left(  \sigma\right)
$. The REE is also asymptotically continuous: given two $d$-dimensional
$\varepsilon$-close states $\left\Vert \rho-\sigma\right\Vert \leq\varepsilon
$, we have $|E_{\mathrm{R}}(\rho)-E_{\mathrm{R}}(\sigma)|\leq4\varepsilon
\log_{2}d+2H_{2}(\varepsilon)$, where $H_{2}$ is the binary Shannon entropy.

\SP{\subsection{Quantum networks: formalism and simulation}}

A quantum network $\mathcal{N}$ can be represented as an undirected finite
graph~\cite{Slepian} $\mathcal{N}=(P,E)$, where $P$ represent the set of
points (or nodes), while $E$ is the set of undirected edges. We assume that
every point $P$\ has a quantum register $\mathbf{p}$, i.e., an ensemble of
quantum systems that are used for quantum communication and local quantum
information processing. Between two points $\mathbf{x}$\ and $\mathbf{y}$,
there is an edge $(\mathbf{x},\mathbf{y})$ if there is a corresponding quantum
channel $\mathcal{E}_{\mathbf{xy}}$. In general, we assume that the channel is
bidirectional, meaning that it can be used in forward direction
$\mathbf{x\rightarrow y}$\ or backward $\mathbf{y}\rightarrow\mathbf{x}$. For
two labeled points $\mathbf{p}_{i}$ and $\mathbf{p}_{j}$, we may also adopt the
simpler notation $\mathcal{E}_{ij}:=\mathcal{E}_{\mathbf{p}_{i}\mathbf{p}_{j}}$. Given two points $\mathbf{a}$\ and $\mathbf{b}$, a cut
$C:\mathbf{a}|\mathbf{b}$ with respect to these points is a bipartition
$(\mathbf{A},\mathbf{B})$ of $P$\ such that $\mathbf{a}\in\mathbf{A}$ and
$\mathbf{b}\in\mathbf{B}$. Given a cut, its cut-set $\tilde{C}$ is defined by
$\tilde{C}=\{(\mathbf{x},\mathbf{y})\in E:\mathbf{x}\in\mathbf{A},\mathbf{y}\in\mathbf{B}\}$, i.e., represents the ensemble of edges across the
bipartition. In general, a cut can be defined between multiple points, i.e.,
we may consider $C:\{\mathbf{a}_{i}\}|\{\mathbf{b}_{j}\mathbf{\}}$ for $i=1,\dots,N$
and $j=1,\dots,M$. This means that the bipartition is such that $\mathbf{a}_{i}\in\mathbf{A}$ and $\mathbf{b}_{j}\in\mathbf{B}$ for any $i$ and $j$.

%\subsection{Simulation of a quantum network}

Given a network $\mathcal{N}$, we may consider its
simulation~\cite{netpaper,longVersion}. This means that, for any edge
$(\mathbf{x},\mathbf{y})$, the quantum channel $\mathcal{E}_{\mathbf{xy}}$ can
be replaced by a simulation $S_{\mathbf{xy}}=(\mathcal{T}_{\mathbf{xy}}%
,\sigma_{\mathbf{xy}})$ where an LOCC $\mathcal{T}_{\mathbf{xy}}$ is applied
to a resource state $\sigma_{\mathbf{xy}}$, so that $\mathcal{E}_{\mathbf{xy}%
}(\rho)=\mathcal{T}_{\mathbf{xy}}(\rho\otimes\sigma_{\mathbf{xy}})$ for any
input state. More generally, this may be an asymptotic simulation
$\mathcal{E}_{\mathbf{xy}}(\rho)=\lim_{\mu}\mathcal{T}_{\mathbf{xy}}^{\mu
}(\rho\otimes\sigma_{\mathbf{xy}}^{\mu})$ with sequences of LOCCs
$\mathcal{T}_{\mathbf{xy}}^{\mu}$ and resource states $\sigma_{\mathbf{xy}%
}^{\mu}$. Therefore, we may define the LOCC\ simulation of the entire network
$S(\mathcal{N})=\{S_{\mathbf{xy}}\}_{(\mathbf{x},\mathbf{y})\in E}$ and a
corresponding resource representation $\sigma(\mathcal{N})=\{\sigma
_{\mathbf{xy}}\}_{(\mathbf{x},\mathbf{y})\in E}$, where $\sigma_{\mathbf{xy}}$
may be asymptotic, i.e., defined by $\sigma_{\mathbf{xy}}=\lim_{\mu}%
\sigma_{\mathbf{xy}}^{\mu}$. In particular, for a network with
teleportation-covariant channels, we may use teleportation LOCCs and the Choi
representation $\sigma(\mathcal{N})=\{\sigma_{\mathcal{E}_{\mathbf{xy}}%
}\}_{(\mathbf{x},\mathbf{y})\in E}$. %We say that the quantum network is distillable when all its channels are distillable.

\section{Multicasts of conferencing keys\label{MainSEC}}

We consider the model of single-message multiple-multicast network in the
quantum setting. Here we have $N$ senders $\{\mathbf{a}_{i}%
\}_{i=1}^{N}=\{\mathbf{a}_{1},\ldots,\mathbf{a}_{i},\ldots,\mathbf{a}_{N}\}$
and $M$ receivers $\{\mathbf{b}_{j}\}_{j=1}^{M}=\{\mathbf{b}%
_{1},\ldots,\mathbf{b}_{j},\ldots,\mathbf{b}_{M}\}$. Each sender
$\mathbf{a}_{i}$ aims at generating the same conferencing secret key $K_{i}$
with all the $M$ receivers. Different senders distribute different keys to the
ensemble of receivers, so that we have a total of $N$ keys. In general we
assume that each point of the network can perform arbitrary LOs on their
registers, assisted by two-way CCs with all the other points of the network.
These adaptive LOCCs can be performed before and after each use of each
channel in the network. We also assume that the global distribution of the $N$
keys is performed assuming a multi-path flooding~\cite{flooding} protocol
$\mathcal{P}$ where each channel of the network is actively exploited by the
parties for each use of the network (see
Refs.~\cite{longVersion,netpaper,netpaper2} for more details on these general
protocols).

More precisely the aim of \SP{the $i$-th sender is to share} copies of a
multipartite private state $\phi_{\mathbf{a}_{i}\{\mathbf{b}_{j}\}}$ with the
destination set \SP{of the $M$ receivers.} This \SP{state} is a
direct generalization of a GHZ state $(\left\vert 0\right\rangle
^{\otimes(M+1)}+\left\vert 1\right\rangle ^{\otimes(M+1)})/\sqrt{2}$\ to
include an additional shield system~\cite{KD1}, \SP{and generates} one private bit
shared between \SP{the sender and all the receivers}. After $n$ uses of the network, the $N$
\SP{senders} and $M$ \SP{receivers} will share a global output state $\rho_{\{\mathbf{a}_{i}\}\mathbf{\{b}_{j}\}}^{n}$ which is $\varepsilon$-close
to the target state
\begin{equation}
\phi:=%
%TCIMACRO{\tbigotimes _{i=1}^{N}}%
%BeginExpansion
{\textstyle\bigotimes_{i=1}^{N}}
%EndExpansion
\phi_{\mathbf{a}_{i}\{\mathbf{b}_{j}\}}^{\otimes nR_{i}^{\varepsilon,n}},
\end{equation}
where $nR_{i}^{\varepsilon,n}$ \SP{is the number of copies distributed by the $i$-th sender}. By taking the limit
of large $n$, small $\varepsilon$, and optimizing over all protocols
$\mathcal{P}$, one defines the capacity region for the achievable key rates
$\{R_{i}\}$. We can then prove \SP{our main} result.

\begin{theorem}
[Single-message multiple multicasts]\label{Theomultimulti}Let us consider a
quantum network $\mathcal{N}=(P,E)$ with resource representation
$\sigma(\mathcal{N})=\{\sigma_{\mathbf{xy}}\}_{(\mathbf{x},\mathbf{y})\in E}$,
which may be a Choi-representation for a teleportation-covariant $\mathcal{N}%
$. Consider the most general multiple-multicast protocol where \SP{the $i$-th} of $N$
senders $\mathbf{\{a}_{i}\}$ distributes an independent key to a destination
set of $M$ receivers $\mathbf{\{b}_{j}\}$ at the rate $R_{i}$. Then, we have
the following outer bound for the capacity region
\begin{equation}
\sum\limits_{i=1}^{N}R_{i}\leq\min_{C:\{\mathbf{a}_{i}\}|\{\mathbf{b}_{j}%
\}}E_{\mathrm{R}}^{\text{\textrm{m}}}(C),\label{setCVB}%
\end{equation}
where $E_{\mathrm{R}}^{\text{\textrm{m}}}(C)$ is the multi-edge flow of
REE\ through cut $C$, defined by%
\begin{equation}
E_{\mathrm{R}}^{\text{\textrm{m}}}(C):=\sum\limits_{(\mathbf{x},\mathbf{y}%
)\in\tilde{C}}E_{\mathrm{R}}(\sigma_{\mathbf{xy}}),
\end{equation}
which is implicitly extended to asymptotic simulations.
\end{theorem}

\textbf{Proof.}~~Consider an arbitrary cut of the type $C:\{\mathbf{a}%
_{i}\}|\{\mathbf{b}_{j}\}$. With respect to this bipartition, we may write the
distillable key $K_{\mathrm{D}}$ of the target state and write
\begin{align}
K_{\mathrm{D}}(\{\mathbf{a}_{i}\}|\{\mathbf{b}_{j}\})_{\phi} &  =n\sum
\limits_{i=1}^{N}R_{i}^{\varepsilon,n} \nonumber\\
&  \overset{\mathrm{(i)}}{\leq}E_{\mathrm{R}}(\{\mathbf{a}_{i}\}|\{\mathbf{b}%
_{j}\})_{\phi} \nonumber \\
&  \overset{\mathrm{(ii)}}{\leq}E_{\mathrm{R}}(\{\mathbf{a}_{i}\}|\{\mathbf{b}%
_{j}\})_{\rho^{n}}+\delta(\varepsilon,d),\label{eq3}%
\end{align}
where we use (i) the fact that the distillable key of a state is upper bounded
by its REE~\cite{KD1}, and (ii) the continuity of the REE\ with respect to the
states $\left\Vert \rho-\phi\right\Vert \leq\varepsilon$, where $\rho
:=\rho_{\{\mathbf{a}_{i}\}\mathbf{\{b}_{j}\}}^{n}$ is the output state and
$\phi$ is the target state. In Eq.~(\ref{eq3}), the error term $\delta
(\varepsilon,d)$ depends on the $\varepsilon$-closeness and the dimension $d$ of
the target private state $\phi$. %In the weak converse limit of large $n$ and
%small $\varepsilon$, one can neglect $\delta(\varepsilon,d)/n$.
%(this is proven in Refs.~\cite{Matthias1a,Matthias2a} for DV systems and in Ref.~\cite{QKDpaper} for both DV and CV systems. See also Ref.~\cite{TQC}).
\SP{More in detail, this error term can be expressed as $\delta(\varepsilon,d)=4\varepsilon \log_2 d+ 2 H_2(\varepsilon)$, where
$H_2(\varepsilon):=-\varepsilon \log_2(\varepsilon)-(1-\varepsilon) \log_2 (1-\varepsilon)$ and the dimension of the private state grows at most
exponentially in $n$, i.e., $d \le 2^{\alpha_n n}$, where $\alpha_n$ tends to a finite constant. This is proven
in Refs.~\cite{Matthias1a,Matthias2a} for discrete-variable systems and Ref.~\cite{QKDpaper} for both discrete- and continuous-variable systems (see also Ref.~\cite{TQC}). Therefore, we may write $\delta(\varepsilon,d) /n \le 4 \varepsilon \alpha_n + 2 H_2(\varepsilon) / n$. By taking the limit for large $n$ and small $\varepsilon$ (weak converse limit), the right hand side goes to zero and we can neglect $\delta(\varepsilon,d)/n$. Therefore, by taking the weak converse limit in Eq.~(\ref{eq3}), we find}
\begin{equation}
\lim_{\varepsilon,n}\sum\limits_{i=1}^{N}R_{i}^{\varepsilon,n}\leq
\underset{n\rightarrow\infty}{\lim}~n^{-1}E_{\mathrm{R}}(\{\mathbf{a}%
_{i}\}|\{\mathbf{b}_{j}\})_{\rho^{n}}.\label{repL}%
\end{equation}

The next ingredient is the simulation of the network. Given a simulation
$S(\mathcal{N})=\{S_{\mathbf{xy}}\}_{(\mathbf{x},\mathbf{y})\in E}$ with
resource representation $\sigma(\mathcal{N})=\{\sigma_{\mathbf{xy}%
}\}_{(\mathbf{x},\mathbf{y})\in E}$ (where we implicitly include asymptotic
states) we may `stretch' any adaptive protocol implemented over the network
using the tools of Refs.~\cite{longVersion,netpaper} and write the output
state in the block form
\begin{equation}
\rho_{\{\mathbf{a}_{i}\}\{\mathbf{b}_{j}\}}^{n}=\bar{\Lambda}\left[
\underset{(\mathbf{x},\mathbf{y})\in E}{%
%TCIMACRO{\tbigotimes }%
%BeginExpansion
{\textstyle\bigotimes}
%EndExpansion
}\sigma_{\mathbf{xy}}^{\otimes n}\right]  ,
\end{equation}
where $\bar{\Lambda}$ is a trace-preserving LOCC. By adopting an arbitrary cut
of the type $C:\{\mathbf{a}_{i}\}|\{\mathbf{b}_{j}\}$, we can reduce this
decomposition into the following%
\begin{equation}
\rho_{\{\mathbf{a}_{i}\}\{\mathbf{b}_{j}\}}^{n}(C)=\bar{\Lambda}_{C}\left[
\underset{(\mathbf{x},\mathbf{y})\in\tilde{C}}{%
%TCIMACRO{\tbigotimes }%
%BeginExpansion
{\textstyle\bigotimes}
%EndExpansion
}\sigma_{\mathbf{xy}}^{\otimes n}\right]  ,\label{toREP}%
\end{equation}
where $\bar{\Lambda}_{C}$ is now local with respect to the bipartition
introduced by the cut $C$. This decomposition is implicitly assumed to be
asymptotic in the presence of asymptotic resource states, in which case it
becomes of the following type
\begin{equation}
\rho_{\{\mathbf{a}_{i}\}\{\mathbf{b}_{j}\}}^{n}(C)=\lim_{\mu}\bar{\Lambda}%
_{C}^{\mu}\left[  \underset{(\mathbf{x},\mathbf{y})\in\tilde{C}}{%
%TCIMACRO{\tbigotimes }%
%BeginExpansion
{\textstyle\bigotimes}
%EndExpansion
}\sigma_{\mathbf{xy}}^{\mu\otimes n}\right]  ,
\end{equation}
for sequences of LOCCs $\bar{\Lambda}_{C}^{\mu}$ and resource states
$\sigma_{\mathbf{xy}}^{\mu}$.

By replacing Eq.~(\ref{toREP}) in Eq.~(\ref{repL}), we may exploit the
monotonicity of the REE\ under trace preserving LOCCs and write%
\begin{equation}
\lim_{\varepsilon,n}\sum\limits_{i=1}^{N}R_{i}^{\varepsilon,n}\leq
E_{\mathrm{R}}^{\text{m}}(C).
\end{equation}
Then, if we minimize over all possible cuts of the type $C:\{\mathbf{a}%
_{i}\}|\{\mathbf{b}_{j}\}$, we may write the following bound for the
asymptotic rates%
\begin{equation}
\sum\limits_{i=1}^{N}R_{i}\leq\min_{C:\{\mathbf{a}_{i}\}|\{\mathbf{b}_{j}%
\}}E_{\mathrm{R}}^{\text{m}}(C),
\end{equation}
which concludes the proof.~$\blacksquare$

\bigskip

Some considerations are in order. First of all, let us note that, for a
distillable network, i.e., a network connected by distillable
channels~\cite{QKDpaper}, such as pure-loss channels, quantum-limited
amplifiers, dephasing and erasure channels, we have a simplification of the
bound. A distillable channel $\mathcal{E}$ is a particular
teleportation-covariant channel whose secret-key capacity $\mathcal{K}$ is
equal to the REE of its Choi matrix, i.e., $\mathcal{K}(\mathcal{E}%
)=E_{\mathrm{R}}(\sigma_{\mathcal{E}})$. Therefore, for a distillable network
with channels $\mathcal{E}_{\mathbf{xy}}$, for any cut $C$, we may write
\begin{align}
E_{\mathrm{R}}^{\text{m}}(C)  & =\sum\limits_{(\mathbf{x},\mathbf{y})\in
\tilde{C}}E_{\mathrm{R}}(\sigma_{\mathcal{E}_{\mathbf{xy}}})\\
& =\sum\limits_{(\mathbf{x},\mathbf{y})\in\tilde{C}}\mathcal{K}(\mathcal{E}%
_{\mathbf{xy}}):=\mathcal{K}^{\text{m}}(C),
\end{align}
where $\mathcal{K}^{\text{m}}(C)$ is the multi-edge secret-key capacity of the
cut $C$~\cite{longVersion,netpaper}.

Then, consider the case of a single \SP{sender} ($N=1$), that we denote by
$\mathbf{a}$. This is the most basic scenario for quantum conferencing in a
multi-hop quantum network. We can see that the bound in Eq.~(\ref{setCVB})
simplifies to
\begin{equation}
R\leq\min_{C:\mathbf{a}|\{\mathbf{b}_{j}\}}E_{\mathrm{R}}^{\text{m}}(C),
\end{equation}
where $R$ is the maximum achievable rate. While this bound is generally large,
there are network configurations where it is sufficiently tight. For instance,
consider the case where \SP{the sender} wants to generate a conferencing key with the
destination set but it is limited to connect to an intermediate router node
$\mathbf{r}$\ via a quantum channel $\mathcal{E}_{\mathbf{ar}}$. Then, it is
immediate to see that the conferencing key must satisfy $R\leq E_{\mathrm{R}%
}(\sigma_{\mathbf{ar}})$, where $\sigma_{\mathbf{ar}}$ is the resource state
associated with the simulation of $\mathcal{E}_{\mathbf{ar}}$. If the channel
is distillable, we then have $R\leq K(\mathcal{E}_{\mathbf{ar}})$. For
instance, if it is a pure-loss channel with transmissivity $\eta$, we find
$R\leq-\log_{2}(1-\eta)$, i.e., the rate of the conferencing key cannot beat
the PLOB bound~\cite{QKDpaper}.

\section{Conclusions\label{SECconclusions}}

In this work, we have studied the ultimate conferencing key rates that are
achievable in a multi-hop quantum communication network. We have considered
the general scenario of single-message multiple-multicast protocols, where $N$
senders communicate with a destination set of $M$ receivers, and each of the
sender aims at generating the same secret key with the entire destination set.
This general case can also be seen as a protocol for the simultaneous
generation of $N$ conferencing keys shared by the $M$ receivers. For $N=1$,
this reduces to the basic configuration considered in the
literature~\cite{conf1,conf2}.

Our results are heavily based on the tools and notions established in
Refs.~\cite{longVersion,netpaper,netpaper2} for quantum networks, and
Ref.~\cite{QKDpaper} for point-to-point communications. In particular, we
exploit the simulation and stretching techniques developed in these previous works to
reduce the most general (adaptive) protocols into a block form, \SP{so that we can derive} a single-letter upper bound  \SP{for the capacity region} in terms of the REE.
Furthermore, our results do not depend on the dimension of the Hilbert space,
in the sense that they apply to quantum conferencing schemes in quantum
networks connected by DV or CV\ quantum channels.

\textbf{Acknowledgments}.~This work has been supported the European Union via
the project `Continuous Variable Quantum Communications' (CiViQ, Grant No. 820466).


\begin{thebibliography}{99}                                                                                               %
\bibitem {NiCh}M. A. Nielsen, and I. L. Chuang, \textit{Quantum computation
and quantum information} (Cambridge University Press, Cambridge, 2000).

\bibitem {BraRMP}S. L. Braunstein and P. van Loock, Rev. Mod. Phys.
\textbf{77}, 513 (2005).

\bibitem {HolevoBOOK}A. Holevo,\textit{ Quantum Systems, Channels,
Information: A Mathematical Introduction} (De Gruyter, Berlin-Boston, 2012).

\bibitem {review}C. Weedbrook \textit{et al.}, Rev. Mod. Phys. \textbf{84},
621 (2012).

\bibitem {first}J. Watrous, \textit{The theory of quantum information}
(Cambridge University Press, Cambridge, 2018).

\bibitem {BB84}C. H. Bennett and G. Brassard,\ Proc. IEEE International Conf.
on Computers, Systems, and Signal Processing, Bangalaore, pp. 175--179 (1984).

\bibitem {Ekert}A. K. Ekert, Phys. Rev. Lett. \textbf{67}, 661-663 (1991).

\bibitem {QKDadvance}S. Pirandola, U. L. Andersen, L. Banchi, M. Berta, D.
Bunan-dar, R. Colbeck, D. Englund, T. Gehring, C. Lupo, C. Ottaviani, J.
Pereira, M. Razavi, J. S. Shaari, M. Tomamichel, V. C. Usenko, G. Vallone, P.
Villoresi, and P. Wallden, ``Advances in quantum cryptography'', preprint
arXiv:1906.01645 (2019).

\bibitem{Frolic} \SP{B. Fröhlich, J. F. Dynes, M. Lucamarini, A. W. Sharpe, Z. Yuan, and A. J. Shields, Nature \textbf{501}, 69-72 (2013).}

\bibitem{satellite1} \SP{G. Vallone, D. Bacco, D. Dequal, S. Gaiarin, V. Luceri, G. Bianco, and P. Villoresi, Phys. Rev. Lett. \textbf{115}, 040502 (2015).}
\bibitem{satellite2} \SP{S.-K. Liao, W.-Q. Cai, J. Handsteiner, B. Liu, J. Yin, L. Zhang, D. Rauch, M. Fink, J.-G. Ren, and W.-Y. Liu et al., Phys. Rev. Lett. \textbf{120}, 030501 (2018).}


\bibitem {Kimble}H. J. Kimble, Nature \textbf{453}, 1023-1030 (2008).

\bibitem {HybridINTERNET}S. Pirandola, and S. L. Braunstein, Nature
\textbf{532}, 169-171 (2016).

\bibitem {Whener}S. Wehner, D. Elkouss, and R. Hanson, Science \textbf{362},
303 (2018).

\bibitem {QKDpaper}S. Pirandola, R. Laurenza, C. Ottaviani and L. Banchi, Nat.
Commun. \textbf{8}, 15043 (2017). See also arXiv:1510.08863 (2015).

\bibitem {ReverseCAP}S. Pirandola, R. Garc\'{\i}a-Patr\'{o}n, S. L.
Braunstein, and S. Lloyd, Phys. Rev. Lett. \textbf{102}, 050503 (2009).

\bibitem {RevCohINFO}R. Garc\'{\i}a-Patr\'{o}n, S. Pirandola, S. Lloyd, and J.
H. Shapiro, Phys. Rev. Lett. \textbf{102}, 210501 (2009).

\bibitem {Marco}M. Lucamarini, Z. L. Yuan, \ J. F. Dynes, and A. J. Shields,
Nature (London) \textbf{557}, 400-403 (2018).


\bibitem{MarcoEXP} \SP{M. Minder, M. Pittaluga, G. L. Roberts, M. Lucamarini, J. F. Dynes, Z. L. Yuan, and A. J. Shields, Nat. Photon. \textbf{13}, 334–338 (2019).}
\bibitem{Liu} \SP{Y. Liu \textit{et al.}, Phys. Rev. Lett. \textbf{123}, 100505 (2019).}


\bibitem {Briegel}H.-J. Briegel, W. D\"{u}r, J. I. Cirac, and P. Zoller, Phys.
Rev. Lett. \textbf{81}, 5932-5935 (1998).

\bibitem {Rep2}W. D\"{u}r, H.-J. Briegel, J. I. Cirac, and P. Zoller, Phys.
Rev. A \textbf{59}, 169 (1999).

\bibitem {Rep3}L. M. Duan, M. D. Lukin, J. I. Cirac, and P. Zoller, Nature
(London) \textbf{414}, 413 (2001).

\bibitem {Slepian}P. Slepian, \textit{Mathematical Foundations of Network
Analysis} (Springer-Verlag, New York, 1968).

\bibitem {Schrijver}A. Schrijver, \textit{Combinatorial Optimization}
(Springer-Verlag, Berlin, 2003).

\bibitem {Gamal}A. El Gamal and Y.-H. Kim, \textit{Network Information
Theory}, (Cambridge Univ. Press, 2011).

\bibitem {Cover}T. M. Cover and J. A. Thomas, \textit{Elements of Information
Theory}, (Wiley, New Jersey, 2006).

\bibitem {netflow}R. K. Ahuja, T. L. Magnanti, and J. B. Orlin,
\textit{Network Flows: Theory, Algorithms and Applications} (Prentice Hall, 1993).

\bibitem {Metro}S. Pirandola, and C. Lupo, Phys. Rev. Lett. \textbf{118},
100502 (2017).

\bibitem {nonPauli}T. P. W. Cope, L. Hetzel, L. Banchi, and S. Pirandola,
Phys. Rev. A \textbf{96}, 022323 (2017).

\bibitem {TQC}S. Pirandola, S. L. Braunstein, R. Laurenza, C. Ottaviani, and
L. Banchi, Quant. Sci. Tech. \textbf{3}, 035009 (2018).

\bibitem {BK2}S. Pirandola, R. Laurenza, and S. L. Braunstein, Eur. Phys. J. D
\textbf{72}, 162 (2018).

\bibitem {Qmetro}R. Laurenza, C. Lupo, G. Spedalieri, S. L. Braunstein, and S.
Pirandola, Quantum Meas. Quantum Metrol. \textbf{5}, 1-12 (2018).

\bibitem {revSENS}S. Pirandola, B. Roy Bardhan, T. Gehring, C. Weedbrook, and
S. Lloyd, Nat. Photon. \textbf{12}, 724-733 (2018).

\bibitem {longVersion}S. Pirandola, \textit{Capacities of repeater-assisted
quantum communications}, arXiv:1601.00966 (2016).

\bibitem {netpaper}S. Pirandola, Commun. Phys. \textbf{2}, 51 (2019).

\bibitem {netpaper2}S. Pirandola, Quantum Sci. Technol. \textbf{4}, 045006 (2019).

\bibitem {telereview}S. Pirandola, J. Eisert, C. Weedbrook, A. Furusawa, and
S. L. Braunstein, Nature Photon. \textbf{9}, 641-652 (2015).

\bibitem {teleBENNETT}C. H. Bennett, G. Brassard, C. Crepeau, R. Jozsa, A.
Peres, and W. K. Wootters, Phys. Rev. Lett. \textbf{70}, 1895-1899 (1993).

\bibitem {Samtele}S. L. Braunstein, and H. J. Kimble,\textit{ }Phys. Rev.
Lett. \textbf{80}, 869--872 (1998).

\bibitem {Samtele2}S. L. Braunstein, G. M. D'Ariano, G. J. Milburn, and M. F.
Sacchi, Phys. Rev. Lett. \textbf{84}, 3486--3489 (2000).

\bibitem {RMPrelent}V. Vedral, Rev. Mod. Phys. \textbf{74}, 197 (2002).

\bibitem {VedFORMm}V. Vedral, M. B. Plenio, M. A. Rippin, and P. L. Knight,
Phys. Rev. Lett. \textbf{78}, 2275-2279 (1997).

\bibitem {Pleniom}V. Vedral, and M. B. Plenio, Phys. Rev. A \textbf{57}, 1619 (1998).

\bibitem {flooding}A. S. Tanenbaum and D. J. Wetherall, \textit{Computer
Networks} (5th Edition, Pearson, 2010).

\bibitem {KD1}K. Horodecki, M. Horodecki, P. Horodecki, and J. Oppenheim,
Phys. Rev. Lett. \textbf{94}, 160502 (2005).

\bibitem {Matthias1a}M. Christiandl, A. Ekert, M. Horodecki, P. Horodecki, J.
Oppenheim, and R. Renner, Lecture Notes in Computer Science \textbf{4392},
456-478 (2007). See also arXiv:quant-ph/0608199v3 for a more extended version.

\bibitem {Matthias2a}M. Christiandl, N. Schuch, and A. Winter, Comm. Math.
Phys. \textbf{311}, 397-422 (2012).

\bibitem {conf1}C. Ottaviani, C. Lupo, R. Laurenza, and S. Pirandola,
Communications Physics \textbf{2}, 118 (2019).

\bibitem {conf2}F. Grasselli, H. Kampermann, and D. Bru\ss,
\SP{New J. Phys. \textbf{21}, 123002 (2019).}

%\textit{Conference key agreement with single-photon interference}, arXiv:1907.10288 (2019).






\end{thebibliography}
\end{document}